\documentclass[preprint,12pt,authoryear]{elsarticle}

\usepackage{amssymb}
\usepackage{amsmath}
\usepackage{makecell} 

\journal{}

\begin{document}

\begin{frontmatter}

\title{Measuring Social Influence with \\
Networked Synthetic Control} 

\author{Ho-Chun Herbert Chang} 

\affiliation{organization={Program in Quantitative Social Science},
            addressline={3 Tuck Mall}, 
            city={Hanover},
            postcode={03755}, 
            state={NH},
            country={USA}}

\begin{abstract}
Measuring social influence is difficult due to the lack of counter-factuals and comparisons. By combining machine modeling and network science, we present general properties of social value, a recent measure for social influence using synthetic control applicable to political behavior. Social value diverges from centrality measures on in that it relies on an external regressor to predict an output variable of interest, generates a synthetic measure of influence, then distributes individual contribution based on a social network. Through theoretical derivations, we show the properties of SV under linear regression with and without interaction, across lattice networks, power-law networks, and random graphs. A reduction in computation can be achieved for any ensemble model. Through simulation, we find that the generalized friendship paradox holds--- that in certain situations, your friends have on average more influence than you do. 
\end{abstract}




\end{frontmatter}



\section{Introduction}\label{sec1}

Understanding the effects of interventions, events, or individuals on social behavior remains a central challenge in computational social science. In the absence of randomized experiments, researchers often rely on observational data to construct counterfactual scenarios—what could have happened—posing what is often a challenge of “speculative fiction.” Traditionally, random field experiments were a principle way to produce direct causal evidence. Beyond randomized experiments, comparative case studies are often used to demonstrate what "could have been," should an event not happened. For instance, studies have investigated the role of Cuban migration in Miami compared to other southern cities~\citep{card1990impact} or examined the effects of terrorist conflict in Basque Country against other Spanish regions~\citep{abadie2003economic}. However, case studies provide a very limited sample size. To address this limitation, synthetic control methods evaluate the effects of policies in absence of such comparisons, by constructing synthetic control groups through weighted combinations~\citep{abadie2015comparative}. For instance, Beckley, Horiuchi, and Miller (2018) constructed a "synthetic" Japan to simulate conditions had the United States not formed an alliance~\citep{beckley2018america}. Growth in the predictive power in machine learning, such as gradient-boosting trees~\citep{natekin2013gradient} and deep learning~\citep{lecun2015deep}, has made synthetic control more feasible and accessible. In particular, advances in large-language models such as ChatGPT have enabled interpretable and linguistically advanced modeling of human subjects~\citep{noh2024llms,park2024generative,li2024frontiers}.

However, for social network analysis, the use of synthetic methods presents another methodological challenge, the computationally extensive nature of pairwise relationships. A network of just 100,000 nodes may have up to 10 billion edges. Additionally, existing quasi-experimental approaches often target nodes and edges, rather than the attributes themselves. Simulation-based method such as "knockout" experiments which simulate changes to network structure should certain nodes be removed. For instance, Chang et al. (2023) knocked out wealth managers in offshore financial networks to demonstrate the critical role of intermediaries, and to simulate sanctions on wealth managers~\citep{chang2023complex}. However, these results are interpreted in terms of the network structure, rather than targeted dependent variables.

Recently, a new framework for calculating social influence~\citep{williams2023social} called \textbf{social value} (SV) has been proposed. By removing social co-variates in predicting the behavior of others, the aggregated change in a person's neighbor's behavior encapsulates social influence. In this sense, it combines pairwise synthetic control with network science to produce a metric for social influence, leveraging the inclusion-exclsuion principle. This method has since been used to disambiguate individual contributions---i.e. their social value---toward political communication~\citep{yang2024quantifying}, online attention, purchasing behavior~\citep{williams2023social}, and ideological change. While the method has received usage, the method's theoretical and mathematical properties have yet to be established analytically. 

This paper generalizes social value by formalizing how three fundamental components interact computation of social influence--- \textbf{network}, \textbf{model}, and \textbf{distribution}. We make three contributions. First, we provides a unified framework for computing social value from synthetic control and networks. Second, we derives analytical results for linear, interaction, and ensemble models across standard graph families. Third, we shows SV’s compatibility with the generalized friendship paradox~\citep{eom2014generalized}.

\section{Related Literature}

\subsection{Centrality Measures and Their Limitations}
Centrality metrics have long been foundational tools in social network analysis (SNA) for assessing the structural importance or influence of nodes~\citep{freeman1979centrality,jackson2012social}. These measures quantify how embedded an individual is within a network, based on graph-theoretic properties such as connectivity, path length, and position in flow or communication structures.

There are a few common measures of centrality. Degree centrality, the most basic measure, counts the number of direct connections a node has. While it serves as a direct measure of popularity, it fails to capture the broader structural context. Closeness centrality measures a node’s average distance to all others in the network, thereby identifying the immediacy of its interaction with other nodes. Betweenness centrality captures the extent to which a node lies on the shortest paths between other nodes, and is typically seen as the measurement for brokerage~\citep{brandes2001faster}. 

Despite their widespread use, these centrality measures are fundamentally structural: they reflect a node’s position in the network topology but not its effect on actual outcomes or behaviors \citep{borgatti2005centrality}. Studies that have introduced these centrality measures are limited~\citep{benyahia2015centrality}. Moreover, these centrality measures are global and computationally intensive; for instance, computing betweenness centrality requires examining all shortest paths in the network, leading to $O(N^3)$ complexity for unweighted graphs \citep{brandes2001faster}, which is often intractable for large online or dynamic networks. 

\subsection{Causal Inference for Social Networks}

In tandem, growth in causal inference has reshaped statistical research especially as applied to the social sciences~\citep{angrist2009mostly,hernan2020causal,imbens2015causal, morgan2014counterfactuals,pearl2000causality,rosenbaum2002observational}. Techniques that have attracted attention include randomization tests~\citep{basu1980randomization}, matching estimators~\citep{abadie2006large,ho2007matching}, regression discontinuity \citep{imbens2008regression,cattaneo2022regression,lee2008randomized}, difference-in-differences \citep{card1994minimum}, and crucially, synthetic control \citep{abadie2010synthetic}. This growth in causal inference has perhaps resisted integration into network science, due to the complex nature of graphs.

Some work has attempted to bridged network analysis and causal inference. Pseudo-causal methods like knock-out experiments have been employed in economic sociology, and successfully measured network fragmentation across the offshore financial network of democratic and autocratic countries~\citep{chang2023complex}. However, since the measured attributes are still global, which resists individual-level analysis. Generative models like exponential random graph models (ERGMs) and stochastic actor-oriented models (SAOMs)~\citep{hunter2008ergm,steglich2010dynamic}, attempt to integrate the individuals within network formation. ERGMs model the probability of observing a given network as a function of its global structure and node-level covariates. SAOMs explicitly model the evolution of networks over time based on actors' decisions. At each small time step, an actor has the probability of changing a tie based on an objective function that reflects network effects and covariates. Both of these models focus on the temporal dimension and distributional outcomes. Network mediation analysis, working in the opposite direction, extends classical mediation frameworks to the context of social networks, and focuses on identifying how structural pathways mediate individual-level variables on outcomes~\citep{duxbury2023problem,duxbury2024micro2,duxbury2023scaling}. For instance, \cite{duxbury2024micro} proposes a framework for micro-to-macro mediation, to model the aggregation of individual-level inputs into emergent macro-level structure. In either of these cases, the principle measurement of these experiments is still the global network structure or macro-behavioral patterns. 

On the other hand, large-scale experimental interventions likely generate the most direct evidence of changes in individual patterns. For instance,  Centola (2010) pioneered large-scale experimental interventions by embedding participants in online networks and demonstrating that behavioral adoption spreads more robustly in clustered-lattice structures than in random networks~\citep{centola2010spread}. Recent collaborations with social media companies provided sufficient variation to track individual effects from randomized experiments, such as key influencers of misinformation~\citep{gonzalez2024diffusion} and the mediating effect of algorithms~\citep{gonzalez2023asymmetric,guess2023reshares,allcott2024effects,nyhan2023like}.

\subsection{Synthetic control and machine learning}

Synthetic control methods provide a principled approach to estimating counterfactual outcomes in comparative case studies where traditional randomized experiments are not feasible~\citep{abadie2010synthetic}.
The promise of machine learning lies in its ability to model human behavior with high accuracy and implicit treatment of non-linear interactions between variables, despite its black box nature.
Tree-based methods like random forests and gradient boosting are popular due to their ability to capture non-linear effects and interaction terms without explicit specification \citep{natekin2013gradient, chen2016xgboost}. Neural network approaches, including deep learning, have also been applied to behavioral sequence modeling, attention prediction, and social media interaction \citep{lecun2015deep}. Moreover, LLMs have recently demonstrated the ability to emulate human subjects for certain tasks in survey research~\cite{aher2023using,tjuatja2024llms}. More complex modeling of their behavior, including simulating counterfactuals, are continuing to emerge.

One criticism of predictive machine learning is their black box nature, and significant work has been done to rectify this. For instance, SHAP values and Bayesian Additive Regression Trees (BART) are used to understand why predictions are made---not just what the outcome is \citep{lundberg2017unified, wager2018estimation}. Based on cooperative game theory, SHAP values represent the marginal contribution of individual features to the underlying model, calculated through testing different powersets of features. These models provide estimates for heterogeneous treatment effects and detect underlying drivers of behavior, even when interactions are complex or latent. In some sense, the goal of synthetic control is agnostic to the interpretability of these models.

While these models have been used successfully for synthetic control in normal regressive settings, their application is scant in conjunction with social networks. Social value is one of the few models that integrate the complexity of machine learning in a scalable way. The goal of this paper is to explore the theoretical properties of social value directly.

\section{Mathematical Model of Social Influence}\label{sec2}

\subsection{Framework of Social Value}

Suppose for every person, you have a target variable $y$ at time $t$. You are trying to predict $y$ based on aggregates from the previous time step, which consists of a collection of covariates, which we divide into asocial $X$ and social variables $S$. We compute SV based on 3 steps:
\begin{enumerate}
    \item \textbf{Model Construction:} Train a model $F(X,S)$ to predict target variable $y$, where $X$ are asocial variables and $S$ is the social variable.
    \item \textbf{Synthetic control: }For every individual $i$, compute:
    \begin{equation}\label{eq:delta-y}
        \Delta y(i) = F(X,S) - F(X,0)
    \end{equation}
    where the social variable is set to $0$. This simulates the amount of $y$ had there been no social influence from other individuals. 
    \item \textbf{Network Distribution:} We then distribute $\Delta y(i)$ back to all neighbors of individual $i$, based on the weight $w_{i,j}$.
\end{enumerate}

The overall formula for social value for $i$ can be summarized as:
\begin{equation} \label{eq:SV}
    SV(i) = \sum_{j \in nei(i)} w_{i,j} \frac{\Delta y(j)}{deg(j)}
\end{equation}
where $j$ iterates over all neighbors of $i$ and $w_{i,j}$ is the network weight between $i$ and $j$. 
In matrix format, then:
\begin{equation}
    \Vec{SV} = \textbf{A} \times \Vec{ \Delta y} \odot \Vec{S^{-1}}
\end{equation}
where $\textbf{A}$ is the weighted adjacency matrix for the underlying network $G$, $\Vec{ \Delta y}$ the simulated difference, and $\Vec{S^{-1}}$ the inverse of the degree for every individual. By definition, $S$ is equal to degree and $\odot$ refers to element-wise multiplication.

Social variables are derived from interaction between other users, such as money exchanged, liking photos, or messaging on an app. For simplicity we assume there is only one social variable or a composite social variable. This is because the social variable is what we use to construct our social network, which we assume to be a single-layer (rather than multiplex). As a result, for every individual, $S$ is also equal to their degree.

Using the asocial and social co-variates, we train a model to predict $y$. These models typically fall into three categories: statistical models (i.e. linear or logistic regression), tree-based models (i.e. XGBoost), and neural nets (i.e. transformers). Regardless of what model, the basic principle is to take in $X$ and $S$ and produce a prediction $y$. In this sense, SV is agnostic to the model used, but the accuracy and quality of the model matters to the quality of synthetic control. A mathematical model captures the way a system interacts with each other, in this case how independent variables interact to produce a dependent one. As such, we can also simulate the behavior of every individual had there been no social interaction in $F(X,0)$, with the difference captured by Eq.~\ref{eq:delta-y}. $\Delta y$ then corresponds to the increase (or decrease) of the target variable due to social influence--- how much $j$ was influenced. 

However, the goal of SV is to quantify the amount of influence an individual generates. The social value of individual $i$ is the sum of how much their neighbors were influenced, weighted by their edge weight. This is captured in Eq.~\ref{eq:SV}. Note, the degree in the denominator is from the neighbor. This is because the marginal increase by neighbor $j$ should be adjusted by the number of neighbors of $j$, rather than $i$. In other words, if all social interaction of $j$ comes from $i$, then $\Delta y (j)$ is attributed to $i$ in full; $i$ could have a large degree by virtue of being popular.

\subsection{Comparisons to other metrics}

In most contexts, social influence is most frequently represented by measures of centrality. Common ones include degree centrality, closeness centrality, betweenness centrality, eigenvector centrality, and PageRank. While these measures of centrality are useful, there are a few meaningful differences. Centrality computations require significant computational and memory resources. Betweenness centrality is an $N^3$ algorithm that requires iteratively computing shortest paths~\citep{brandes2001faster}; PageRank and Eigenvector centrality require repeated matrix multiplication in the process of expectation maximization~\citep{bonacich2007some}. Second, these measures are not directly interpretable--- every quanta attributed to an individuals SV corresponds to a dollar spent or minute online. 

The other means of computing the "value" of an individual is using Shapley Values. Originating from game theory, the goal of Shapley values is to understand how much a player $i$ contributes to the outcome of a coalition~\citep{winter2002shapley}. The idea is to compute every power set for which $i$ is present, minus ones for which they are not. SV diverges from Shapley values in two ways. First, Shapley values typically require empirical situations where the player is not present, rather than simulation via a model. It also is extremely computationally extensive as it considers every power set for which an individual is not present; SV instead exploits the underlying network structure. 

In sum, social value can be seen as a hybrid of centrality and Shapley-values. Social value considers a localized power set to aggregate marginal contributions (i.e. influence), while exploiting a weighted network structure. It only considers the immediate neighborhood compared to eigenvector centrality and PageRank, but emerges as more directly interpretable especially relative to independent variables.

\section{Theoretical Properties}

While boasting high accuracy, machine learning models lacks a theoretical grounding shared by probability and traditional statistics. Our goal here is to investigate the theoretical and mathematical properties that are built on three elements: canonical regressive models, distributions, and network structures.

\subsection{Models of Regression}
\subsubsection{Linear Models}

Under linear regression, the model assumes the underlying co-variates and dependent variable hold the following relationship:
$$
    y = \sum_{r=0}^k \beta_r x_r  + \beta_0  =  \Vec{ \beta } \cdot \Vec{ X } + \beta_0 
$$
Assuming social covariate $S$ is on of the independent variable, we can then separate the equation as $y = \Vec{ \beta } \cdot \Vec{ X } + \beta_s S + \beta_0$, assuming a total of $k$ covariates. 
The general derivation of social value under an ordinary linear model can be shown, first with the synthetic control step in Eq~\ref{eq:linear-synthetic}:
\begin{equation} \label{eq:linear-synthetic}
    \begin{aligned}
        \Delta y (j)    &= F(X,s) - F(X,0) \\
                    &=  \Vec{ \beta } \cdot \Vec{ X } + \beta_S S_j + \beta_0 - (\Vec{ \beta } \cdot \Vec{ X } + \beta_S \cdot 0 + \beta_0 ) 
                    &= \beta_S S_j
    \end{aligned}
\end{equation}
All terms except for $S$ are canceled out in the linear model. The formula for SV is thus Eq~\ref{eq:linear-sv}:
\begin{equation} \label{eq:linear-sv}
\begin{aligned}
    SV(i)   &= \sum_{j \in nei(i)} w_{i,j} \frac{\Delta y(j)}{deg(j)} \\
            &= \sum_{j \in nei(i)} w_{i,j} \frac{\beta_S S_j}{S_j} 
            &= \beta_S \sum_{j \in nei(i)} w_{i,j} \\
            &= \beta_S S_i
\end{aligned}
\end{equation}

Thus, under the linear model, social value is the coefficient of the social covariate under the model, multiplied by the degree of the individual. As such, the distribution of SV is straight-forward, with the mean as $\mu_{SV} = \beta_S \mu_S$ and the variance as $Var(SV)^2 = \beta_S^2 Var(S)$.

\subsubsection{Linear Regression with Interaction}

What if there are interactions between the independent variables? In other words, what if there are non-linear relations within the weighted sum? Our strategy for analyzing linear regression with interaction is similar to the base case. The general formula is:
$$
    y = \sum_{r=0}^k \beta_r x_r  + \sum_{r=0,q=0} \beta_{r,q} x_r x_q + \beta_0 
$$
where we have additional interaction terms. We can then cancel out terms as follows:
\begin{equation}
\begin{aligned}
    \Delta y(i) &= F(X,S) - F(X,0) \\
                &= \Big( \sum_{r=0}^k \beta_r x_r  + \sum_{r=0,q=0} \beta_{r,q} x_r x_q 
                + \sum_{r=0} \beta_{r,s} S x_r + \beta_s S + \beta_0 \Big) - \\
                & \qquad \Big( \sum_{r=0}^k \beta_r x_r  + \sum_{r=0,q=0} \beta_{r,q} x_r x_q + \beta_0  \Big) \\
                &= \sum_{r=0} \beta_{r,s} S x_r + \beta_s S = S \Big( \sum_{r=0} \beta_{r,s} x_r + \beta_s \Big) \\
                &= S ( \Vec{\beta}_s \cdot \Vec{x} )
\end{aligned}
\end{equation}
where $\beta_s$ are all the coefficients that involve $S$. In other words, the simulated difference of based on the interaction model are only the terms that contain the social co-variate $S$. Thus, the SV of $i$ is:
\begin{equation} \label{eq:SV-Interaction}
    \begin{aligned}
        SV(i)   &= \sum_{j \in nei(i)} w_{i,j} \frac{\Delta y(j)}{deg(j)} 
                = \sum_{j \in nei(i)} w_{i,j} \frac{S_j \Vec{\beta}_s \cdot \Vec{x}_j  }{S_j}  \\
                &= \sum_{j \in nei(i)} w_{i,j} \Vec{\beta}_s \cdot \Vec{x}_j  
                 = \Vec{w}^T \times \textbf{X} \times \Vec{\beta}_s 
    \end{aligned}
\end{equation}
where $\Vec{w}_i^T$ are the edge weights of $i$ and $X$ the data aggregates of all individuals. Note, if there is no edge between $i$ and $j$, then the weight is 0 and $x_j$ has no bearing on $SV(i)$.

Eq.~\ref{eq:SV-Interaction} contains several interesting properties. First, note that similar to the base case, $S_j$ cancels out. This means SV of $i$ only depends on the asocial covariates of $i$'s neighbors. However, the weights do matter; if you are connected to those who predict highly via their asocial covariates, you will have a high social value. Second, $\Vec{w}^T$ effectively weights the rows of $\textbf{X}$ and $\Vec{\beta}$ weights the columns. As such, we could take the outer product of $\Vec{w}^T \Vec{\beta}$, then multiply this element-wise with $\textbf{X}$, then sum to produce SV as well.

To investigate mean behavior, we consider the expected value of SV. The main strategy is to note that the expected value of each distribution is the mean:
\begin{equation}
    \begin{aligned}
        E\big[ SV(i) \big] &= \sum_{j \in nei(i)} w_{i,j} \Vec{\beta}_s \cdot E \big[\Vec{x}_j \big] 
        = \sum_{j \in nei(i)} w_{i,j} \Vec{\beta}_s \cdot \Vec{\mu_X} \\
        &= S_i \Vec{\beta}_s \cdot \Vec{\mu_X} 
    \end{aligned}
\end{equation}
In other words, SV can be approximated in a similar fashion as base linear regression, but instead of $S_i \beta_S$, it is the multiplication of $S_i$ and the asocial distribution means weighted by the model coefficients.

\subsubsection{Ensemble models}
As we venture into more complex models, generalities are harder to come by due to the complex interaction of variables. However, the general strategy for model simplification still holds. For instance, modern decision tree models such as random forest and gradient-boosting trees are immensely popular and in practice, and what are to build the social value model~\citep{biau2016random,chen2016xgboost}. More generally, they are a class of ensemble model of the form:
$$
F(x)=\sum _{m=1}^{M}\gamma _{m}h_{m}(x)+{\mbox{const}}
$$
which denotes a collection of $M$ models based on base model $h$, then weighted linearly by $\gamma$. For gradient-boosting trees, $h$ is a base decision tree. Similar to linear models, variable $S$ only interacts with other variables in a base model $h_m$ if it is included in the decision process. In other words, all trees such that $h(X,S) = h(X)$ do not feature variable interaction with $S$. These trees will be canceled out similar to the interaction terms that do not include $S$.

Formally, we split the ensemble model $F(X,S)$ into base models that include $S$ and do not include $S$. The marginal contribution can then be expressed as:
$$
\begin{aligned}
    \Delta y(i) &= F(X,S) - F(X,0) \\
    & =
    \sum _{m; S \notin h_m}^M \gamma _{m}h_{m}(X) + {\mbox{const}_1} + 
    \sum _{m; S \in h_m}^M \gamma _{m}h_{m}(X,S) +{\mbox{const}_2} \\
    & \qquad -  
    \sum _{m; S \notin h_m}^M \gamma _{m}h_{m}(X) + {\mbox{const}_1}  -
    \sum _{m; S \in h_m}^M \gamma _{m}h_{m}(X,0) +{\mbox{const}_2} \\
    &= \sum _{m; S \in h_m}^M \gamma _{m} \Big( h_{m}(X,S) - h_{m}(X,0) \Big)
\end{aligned}
$$
The $\Delta y$ used in social value computations is the difference of sub-trees that use $S$, weighted by $\gamma$.

\subsection{Network Considerations}

We consider the behavior on three types of graphs. First, we consider \textbf{lattice graphs} due to the regularity of their degree ($Var(S) = 0$). Second, we consider \textbf{scale-free graphs}, as systems with human behavior often follow some form of long-tail distribution. Third, we consider \textbf{random graphs}, for which connections between nodes are produced strictly at random.

\textbf{Lattice} graphs, frequently known as grid graphs, are characterized by regularity in degree. They are defined by a number $k$, which indicate the number of adjacent nodes. As such, the $E[S] = k$ and $Var(S) = 0$. 
\textbf{Scale-free networks} are characterized by degree distributions that follow a long-tail, power-law distribution:
$$
P(k) \sim k^{-\gamma}
$$
One explanation for the generation of these networks is "preferential attachment," where when nodes enter a network they probabilistically choose their connections based on the degree of existing nodes. This produces the phenomenon where the "richer get richer." The Barabasi-Albert Model captures this process and is parameterized by the total number of nodes $n$, and the number of edges each new node attaches to $m$. As such, the average degree is $2m$. Due to the power-law nature of the degree distribution, the variance is undefined.
\textbf{Random graphs} are graphs whose degree distribution is determined in some random, probabilistic way. The Erdos-Renyi (ER) random graphs parameterizes these graphs through two parameters: the total number of nodes $n$ and the probability of connecting to another node $p$. The expected degree is thus $np$ and the degree variance $np(1-p)$.


\begin{table}[!htb]
\centering
\begin{tabular}{|l|l|l|l|}
\hline
\multicolumn{4}{|c|}{\textbf{Expected Social Value}} \\ \hline
\multicolumn{1}{|l|}{} &
\multicolumn{1}{c|}{\textbf{Lattice (K)}} &
\multicolumn{1}{c|}{\makecell{\textbf{Barabasi-Albert}\\\textbf{(n,m)}}} &
\makecell{\textbf{Erdos-Renyi}\\\textbf{(n,p)}} \\ \hline
\textbf{Linear Model} &
$k \beta_S$ &
$2 m \beta_S$ &
$n p \beta_S$ \\ \hline
\makecell{\textbf{Linear Model} \\ \textbf{+ Interaction}} &
$k \Vec{\beta} \Vec{\mu_X} + k \beta_S$ &
$2 m \Vec{\beta} \Vec{\mu_X} + 2 m \beta_S$ &
$n p \Vec{\beta} \Vec{\mu_X} + n p \beta_S$ \\ \hline
\end{tabular}
\caption{Expected mean social value based on graph parametrization.}
\end{table}

\subsection{Distribution Considerations}
For many of these results, we have an implicit assumption that there is no correlation between social co-variates $S$ and asocial co-variates $X$. However, many of these variables may be correlated in reality--- for instance, popularity is often correlated to income. As such, we also need to consider situations where $X$ and $S$ have a nonzero correlation. 

For simplicity, we consider just one asocial variable $x$. Assuming the variables have been normalized, then we state $x_j = c_j S_j + f$. Then SV can be characterized as:
\begin{equation} \label{eq:SV-correlation}
    \begin{aligned}
        SV(i)   &= \sum_{j \in nei(i)} w_{i,j} b_x x_j = \sum_{j \in nei(i)} w_{i,j} b_x ( c_j S_j +f)
    \end{aligned}
\end{equation}
The question then becomes, does $w_{i,j}$ correlate with $S_j$? That is, are people whose friends are more popular also more popular? 

To answer this, we consider the Friendship Paradox, where it was mathematically proven that on average, your friends are more popular than you are. More recently, the generalized friendship paradox (GFP) demonstrates that so long as a node attribute correlates with degree, then for that attribute the friendship paradox will also hold~\citep{eom2014generalized}. In other words, if $\beta_x$ and $c$ are both positive or both negative, then $SV$ on average will increase linearly (by $\mu_x$) and on average your friends will have a higher SV; if the signs of $\beta_x$ and $c$, then your friends will have less SV.

\section{Empirical and Simulated Results}

Next, we validate these properties of SV under the linear and interaction models. For the lattice, scale free, and random graph, we initialize graphs of size $n=10,000$ with parameters such that the average degree is 4. This is the default for the lattice graph; for the scale free we set $m=2$, or for every new node added it makes 2 connections; for the random graph we choose $p=\frac{4}{10000}$ such that $np=4$. Note, since $np > 1$, there is a high probability of a large connected components. For simplicity of simulation, we only consider one asocial co-variate $X \sim N(2,4)$. We choose 2 as the mean as it is the first nonnegative integer that does not have multiplicative properties that may lead to some form of cancellation (0 is the origin and 1 is the identity); 4 as it is the expected number of edges.

\subsection{Empirical outcomes on three networks}

\begin{figure}[!htb]
    \centering
    \includegraphics[width=1.0\linewidth]{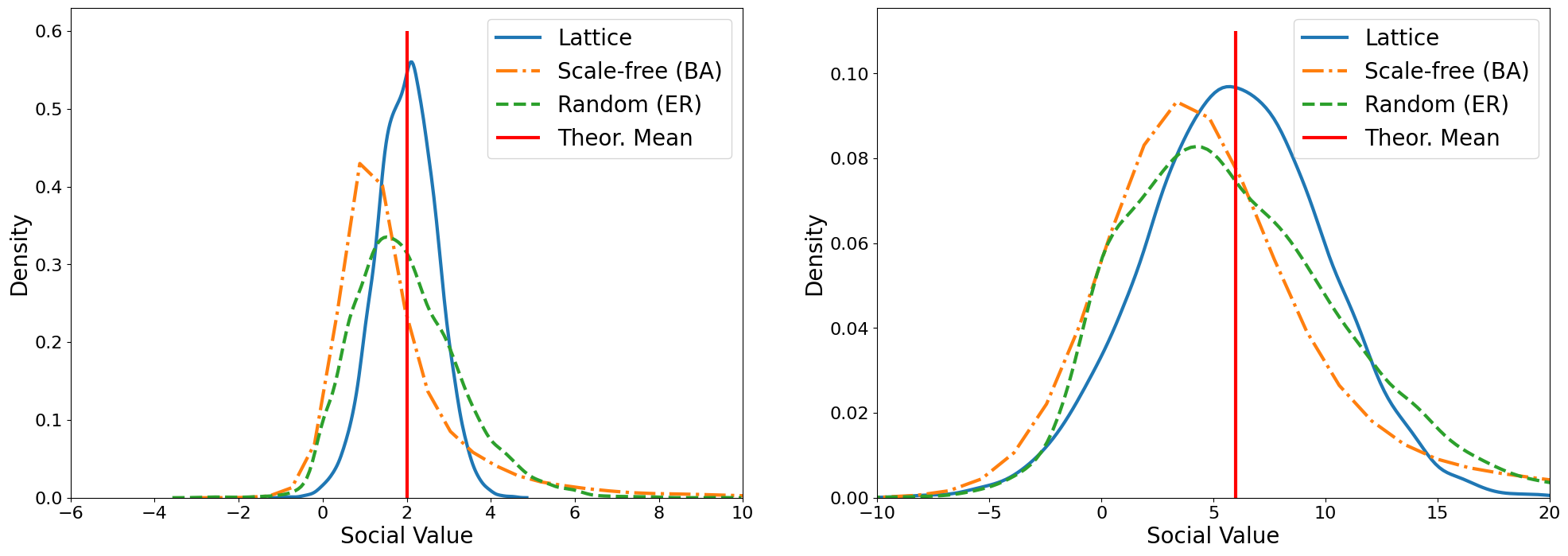}
    \caption{Social Value distributions under a) linear models and b) linear models with interaction.}
    \label{fig:distributions}
\end{figure}

Figure~\ref{fig:distributions} shows the distribution of social value under these assumptions, as a validation to our theoretical expectations. Figure~\ref{fig:distributions}a) shows the distribution under the linear model, with the average degree for all three cohering with the expectation. Moreover, what's interesting is because it only scales by $\beta_S$, the distribution of SV under the linear model is the same as the degree distribution itself. For the lattice, this is symmetric centered around 2, with some spread due to the Gaussian noise in our model. The ER graph resembles the Poisson distribution--- the expected outcome of the random graph degree distribution; the scale-free graph appears to be a powerlaw. In other words, under the linear model we expect the SV distribution to be the degree distribution scaled by $\beta_S^2$ (variance under constant multiplication).

Figure~\ref{fig:distributions}b) in contrast shows the distribution under the linear model with interaction. The theoretical mean is still consistent, at 6. However, the distributions appear much more rounded, due to the mixing with the normal distribution $X$. The tail behavior is preserved---the scale free graph has the longest tail, followed by the random graph, then the lattice.

\subsection{Your friends have more Social Value than you}

Previously, we assume that the node degree $S$ and our node attributes $X$ are independent. What happens when they are correlated? Based on our derivation in Eq.~\ref{eq:SV-correlation}, we expect that when $X$ correlates positively with $S$, and their interaction is positive ($\beta_{X,S} > 0$), then social value should increase. 

\begin{figure}[!htb]
    \centering
    \includegraphics[width=0.85\linewidth]{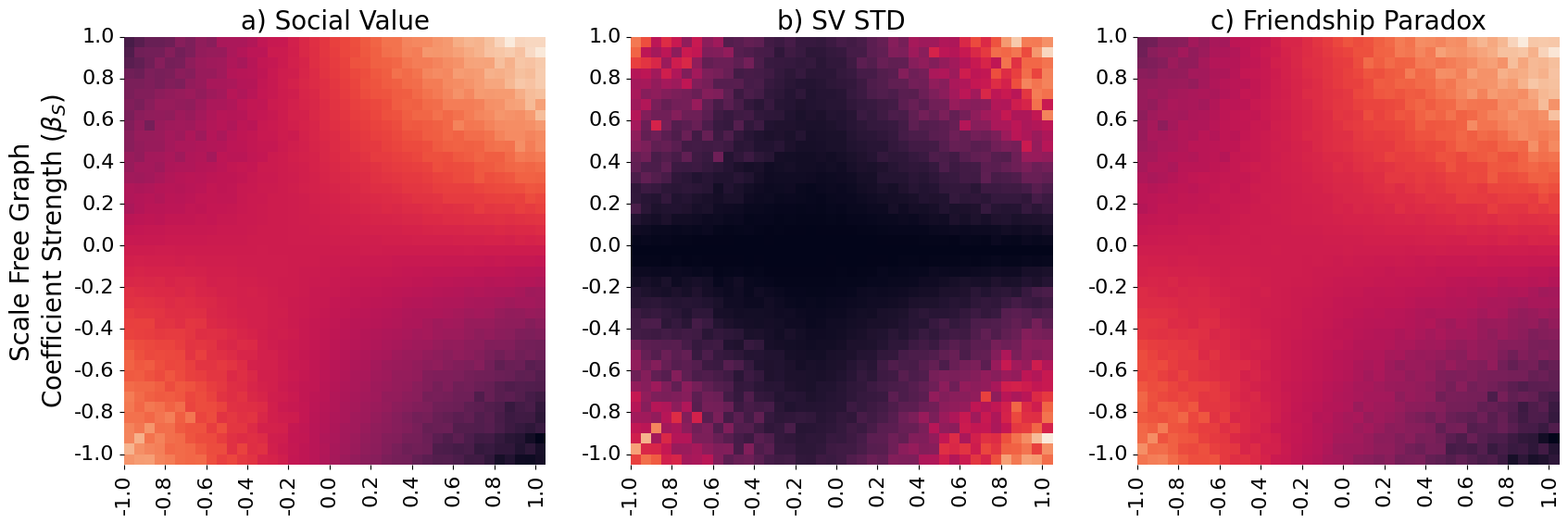}
    \includegraphics[width=0.85\linewidth]{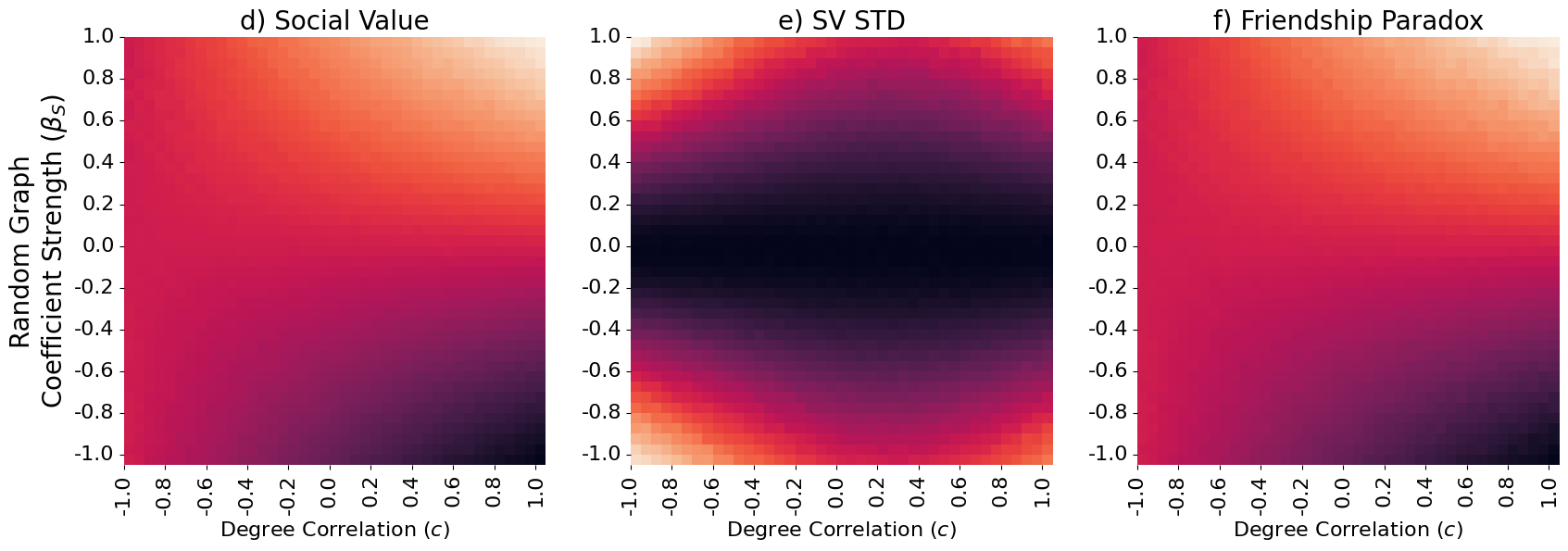}
    \caption{Heat map for social value by varying the coefficient strength $\beta_S$ and correlation $c$ between asocial co-variates and degree. Results are shown for the scale-free graph on the top row, with the mean SV (a), standard deviation of SV (b), and the friendship paradox generalized to SV in (c); for the random graph, the mean SV (d), standard deviation of SV (e), and GFP of SV in (f). }
    \label{fig:paradox}
\end{figure}

Figure~\ref{fig:paradox}a) shows that for BA-graphs when their correlation $c$ increases  and their interaction positively increases, then social value increases. Similarly, when $c$ and $\beta_{X,S}$ both decrease, then SV also increases. In contrast, SV decreases when $c$ and $\beta_{X,S}$ have a different sign. What's interesting is the effect of this interaction is not as pronounced for ER-graphs (d) compared to BA-graphs (a). While change in coefficient strength (in the same range) creates big change in overall SV, this is not the case. Indeed, when we consider Figure~\ref{fig:paradox}b) and e), we find that while the four corners for the scale-free graph all yield high variance, there is only variance across the y-axis for the random graph. This is likely because the degree distribution of the random graph does not vary large enough to generate big changes in SV. 

Lastly, Figure~\ref{fig:paradox}c) and f) show the average difference in SV between a node and its neighbors:
$$
GFP = \frac{1}{N} \sum_{i=0}^N \frac{1}{|nei(i)|} \sum_{j\in nei(i)} SV(j) - SV(i)
$$
The results largely mirror the results of their respective SV means, though on a larger scale. When popularity positively correlates with the asocial co-variate(s), and interacts positively, then your friends have on average more SV; when popularity is negatively correlated with asocial co-variates and interacts negatively, your friends also have on average more SV. SV tends to become a more sensitive measure when dealing with social networks with heavy tail distributions.

\section{Conclusion}
In this paper, we present the general properties of social value, which combines an external regressor for synthetic control and an underlying network to attribute individual influence. The measure diverges from canonical network measures that only consider structure, and game-theoretic measures like Shapley value that considers sets of individuals without exploiting underlying networks.

Theoretical results show that under the linear model, social value varies by it's regression co-efficient $\beta_S$; under the linear model with interaction, SV depends on all terms that have an interaction with $S$. These derivations depend on the cancellation of non-interaction terms. This framework can be extended to any ensemble model given the linear combination of base models, such as random forest, by canceling out base models that do not include an interaction with $S$. Through the empirical results, I also show the general friendship paradox also holds for social value, but it's direction depends on the correlation between asocial and social variables, and the direction of interaction. There is also evidence that under random graphs, SV is less sensitive to interaction strength. The interaction between network configuration and the distribution of social value is worthy of further study.

\bibliographystyle{elsarticle-harv}
\bibliography{cas-refs}

\end{document}